\documentclass[%
superscriptaddress,
 aps,prd,
 twocolumn,
 tightenlines,floatfix
]{revtex4-1}

\usepackage{graphicx}
\usepackage{dcolumn}
\usepackage{bm}
\usepackage{amsmath,amssymb}
\usepackage{natbib}
\usepackage{color,units}
\usepackage{float}
\usepackage{subfigure}
\usepackage[caption=false]{subfig}
\graphicspath{ {./Plots/} }
\usepackage{aas_macros}

\definecolor{darkgreen}{rgb}{0.0, 0.5, 0.0}

\def\fgw{f_\text{gw}}
\def\fgwo{f_{\text{gw},0}}

\begin{document}


\title{X-ray guided gravitational-wave search for binary neutron star merger remnants\\}

\author{Nikhil Sarin}
 \email{nikhil.sarin@monash.edu}
\author{Paul D. Lasky}%
\author{Letizia Sammut}
\author{Greg Ashton}
\affiliation{Monash Centre for Astrophysics, School of Physics and Astronomy, Monash University, VIC 3800, Australia \\
OzGrav: The ARC Centre of Excellence for Gravitational-wave Discovery, Clayton, Victoria 3800, Australia}%

\date{\today}

\begin{abstract}
X-ray observations of some short gamma-ray bursts indicate that a long-lived neutron star can form as a remnant of a binary neutron star merger.
We develop a gravitational-wave detection pipeline for a long-lived binary neutron star merger remnant guided by these counterpart electromagnetic observations.
We determine the distance out to which a gravitational-wave signal can be detected with Advanced LIGO at design sensitivity and the Einstein Telescope using this method, guided by X-ray data from GRB140903A as an example. 
Such gravitational waves can in principle be detected out to $\sim$ 20 Mpc for Advanced LIGO and $\sim$ 450 Mpc for the Einstein Telescope assuming a fiducial ellipticity of $10^{-2}$.  
However, in practice we can rule out such high values of the ellipticity as the total energy emitted in gravitational waves would be greater than the total rotational energy budget of the system.  
We show how these observations can be used to place upper limits on the ellipticity using these energy considerations.  For GRB140903A, the upper limit on the ellipticity is $10^{-3}$, which lowers the detectable distance to $\sim$ 2 Mpc and $\sim$ 45 Mpc for Advanced LIGO and the Einstein Telescope, respectively.
\end{abstract}

\pacs{Valid PACS appear here}
\maketitle

\section{\label{sec:intro}Introduction}
The era of gravitational-wave multi-messenger astrophysics has begun. On 17th August 2017, the Advanced Laser Interferometer Gravitational-wave Observatory (aLIGO) \citep{aligo} and Advanced Virgo \citep{Virgo} made the first gravitational-wave observation of a binary neutron star merger, known as GW170817 \cite{GW170817}. This event was also detected 1.74 seconds later as a short gamma-ray burst (SGRB) by the Fermi and Integral telescopes \cite{GW170817A_GRB}, confirming that binary neutron star mergers can be the progenitors of SGRBs. There are competing hypotheses for the fate of the post-merger remnant. Some analyses of the electromagnetic observations support a hypermassive neutron star that collapsed to form a black hole in $\lesssim 1$s
\citep{Metzger2018, Pooley2017, Margalit2017}. Others support the formation of a stable, rapidly spinning, long-lived magnetar \citep{Yu2018}.

In either case, a short- or long-lived post-merger remnant emits gravitational waves. The detection of such gravitational waves will have significant implications for the understanding of neutron-star physics including the nuclear equation of state. 
A search for short and intermediate duration gravitational-wave signals from a post-merger remnant of GW170817 did not return a significant result \cite{postmerger2017}. This lack of detection was expected given theoretical models \cite{DallOsso2014,Lasky2016a,Doneva2015} and current aLIGO sensitivity. However, the proximity of GW170817, in conjunction with planned upgrades to aLIGO and Virgo sensitivity \citep{observing_aligo} and improved algorithms, suggests, that we may be able to detect post-merger gravitational waves from GW170817-like remnants in the future.

In general, the merger of two neutron stars could result in four different outcomes, which depend on the mass and spin of the remnant and the equation of state - a stable neutron star, a supramassive neutron star, a hyper massive neutron star or the direct collapse to a black hole. A supramassive neutron star is initially supported against gravitational collapse by rigid-body rotation but will collapse to form a black hole on timescales of $10 \textrm{s} - 10^4 \textrm{s}$ \citep{Ravi2014}. A hypermassive neutron star is supported against gravitational collapse through differential rotation but collapses to a black hole in $\leq 1$s (see \citet{Baiotti2017} for a recent review). 

In this paper, we focus on the scenario where a neutron star merger produces a supramassive or stable neutron star remnant. This rapidly spinning star spins down through a combination of electromagnetic and gravitational-wave radiation. 
The latter is likely produced by the non-zero stellar ellipticity in conjunction with the spin-flip instability \citep{Cutler2002, Lasky2016a}, unstable r-modes \citep{ANDERSSON2001, Owen1998} or the secular Chandrasekhar-Friedmann-Schutz bar-mode instability \citep{Lai1995,Shapiro1998,Coyne2016,Shibata2000,Corsi2009,Doneva2015}.

The extended X-ray emission of many SGRBs has been observed by satellites such as \textit{Swift} and Chandra, and used to determine parameters of the neutron star remnant \cite[e.g.,][]{Rowlinson2013, Lu2015, LaskyLeris2017}. 
\citeauthor{Rowlinson2010}\citep{Rowlinson2010,Rowlinson2013} showed that models of magnetic dipole radiation from spinning down millisecond magnetars \cite{Zhang2001, Dai1998} agree with X-ray afterglow observations of several SGRBs. GRB170817A had an extended emission of a different structure \citep[e.g.,][]{JJRuan2018,Troja2017}.

In this paper, we present a method to search for gravitational waves from a long-lived post-merger neutron star remnant. In Sec. \ref{sec:waveform} we derive a model for the gravitational waves emitted from a rapidly spinning down millisecond magnetar while also describing the parameters and the parameter space. 
In Sec. \ref{sec:GRBafterglow} we discuss how we can utilize observations of X-ray afterglows from SGRBs to constrain parameters and run a targeted gravitational-wave search. 
We continue in Sec. \ref{sec:pipeline} with a discussion of the detection statistics for our pipeline and conclude in Sec. \ref{sec:conclusion} with a brief discussion on the extensions that will improve the analysis and physical theory.
\section{Gravitational waveform from millisecond magnetars}\label{sec:waveform}
A long-lived post-merger remnant spins down due to electromagnetic and gravitational-wave radiation. 
We start with the general torque equation. 
\begin{equation}\label{Ch2: Eq. Torque}
\dot{\Omega} = -k\Omega^{n},
\end{equation}
where $\Omega$ and $\dot{\Omega}$ are the star's angular frequency and its time derivative, respectively, $k$ is a constant of proportionality, and $n$ is the braking index. The gravitational-wave frequency is a function of the star's spin frequency. Throughout this work, we assume the gravitational waves are emitted at twice the star's spin frequency, which is true for an orthogonal rotator. The following equations are therefore not valid for gravitational waves from $r$-mode emission; we discuss generalizations of our model in Sec. ~\ref{sec:conclusion}.

The braking index is related to the emission mechanism; $n = 3$ implies that the neutron star is spun down only through a dipole magnetic field in vacuum \citep{1983Shapiro}, while $n = 5$ implies that the neutron star is spun down through gravitational-wave radiation \citep{Yue2006,Bonazzola1996}. A braking index of $n = 7$ is conventionally associated with spin down through unstable $r$ modes \citep[e.g.][]{Owen1998}, although the true value can be less for different saturation mechanisms \citep{Alford2014a,Alford2014b}.
Inference of the braking index for two millisecond magnetars born in SGRBs give $n = 2.9 \pm 0.1$ and $2.6 \pm 0.1$ for GRB130603B and GRB140903A, respectively \cite{LaskyLeris2017}.

Integrating Eq. (\ref{Ch2: Eq. Torque}) and solving for the gravitational-wave frequency gives the gravitational-wave frequency evolution
\begin{equation}\label{Ch2: Eq.fgw(t)}
\fgw(t) = \fgwo\left(1+ \frac{t}{\tau}\right)^{\frac{1}{1 - n}},
\end{equation}
where
\begin{equation}
\tau = \frac{\left(\fgwo\pi\right)^{1 - n}}{-k(1-n)},
\end{equation}
is the spin-down timescale and $\fgwo$ is the gravitational-wave frequency at $ t = 0$. 

The dimensionless gravitational-wave strain amplitude for a non-axisymmetric, rotating body obeying Eq. (\ref{Ch2: Eq. Torque}) is given by 
\begin{equation}\label{Ch2: Eq.strainamp(t)}
h_0(t) = \frac{4\pi^2 G I_{zz}}{c^4} \frac{\epsilon}{d} \fgwo^2\left(1+ \frac{t}{\tau}\right)^{\frac{2}{1 - n}}.
\end{equation}
Here, $I_{zz}$ is the principle moment of inertia, $\epsilon$ is the ellipticity of the rotating body, $d$ is the distance to the source,  $G$ is the gravitational constant, and $c$ is the speed of light. 
The gravitational-wave strain at a detector $h(t)$ is a combination of the $h_+$ and $h_{\times}$ polarisations,
\begin{equation}\label{Ch2: Eq. ht}
h(t) = h_{0}(t)\left[F_{+}\frac{1+\cos^{2}(\iota)}{2}\cos\Phi(t) + F_{\times}\cos(\iota)\sin\Phi(t)\right],
\end{equation}
where, $\iota$ is the inclination angle, and
\begin{equation}\label{Ch2: Eq. phi(t)}
\Phi(t) = \Phi_{0} + 2\pi\int_{0}^{t}dt'\fgw(t'),
\end{equation}
is the phase, with $\Phi_0 = \Phi(t = 0)$. In Eq. (\ref{Ch2: Eq. ht}), $F_+$ and $F_{\times}$ are the antenna pattern functions \citep{Jaranowski1998} for each of the polarisations. 
In reality, $F_+$ and $F_{\times}$ are functions of time. 
In this work, we have ignored this complication and assumed constant $F_+$ and $F_{\times}$ which we determine using the sky location of GRB140903A. 
This does not significantly affect our quantitative results, although it will need to be included when the full pipeline is developed to search for gravitational waves.

Substituting the gravitational-wave frequency evolution from Eq. (\ref{Ch2: Eq.fgw(t)}) into Eq. (\ref{Ch2: Eq. phi(t)}) gives
\begin{equation}\label{Ch2: Eq. Phi}
\Phi(t) = \Phi_0 + 2\pi\tau \fgwo\left(\frac{1 - n}{2 - n}\right)\left[\left(1 + \frac{t}{\tau}\right)^{\frac{2-n}{1 - n}} - 1\right].
\end{equation}
The full waveform model for a rapidly rotating neutron star spinning down due to gravitational wave radiation with an arbitrary braking index consists of Eq. (\ref{Ch2: Eq.strainamp(t)}), (\ref{Ch2: Eq. ht}), and (\ref{Ch2: Eq. Phi}). We refer to this waveform model as the magnetar waveform model, which is parameterized by the initial gravitational-wave frequency $f_{gw,0}$, the spin-down timescale $\tau$, braking index $n$, inclination $\iota$, initial phase $\Phi_0$ and scaling parameters $I_{zz}$, $\epsilon$, $d$.  

In the following, we develop an algorithm for a matched-filter search for gravitational waves using the magnetar waveform model. We construct a template bank by choosing physical parameters for $\fgwo$, $\tau$, $n$, $\iota$, and $\Phi_0$ from a prior. 
We quantify in Sec. \ref{sec:pipeline} that a template bank constructed from physically motivated but unconstrained priors is computationally expensive for detecting gravitational waves, but these priors can be further constrained using X-ray afterglow observations which reduce the computational cost of searches and increase the sensitivity. 
The scaling parameters do not require priors as they only affect the amplitude of the gravitational wave which is normalised in a matched-filter search. Throughout this work, we assume a fiducial moment of inertia, $I_{zz} = 10^{45}$ $\text{g cm}^{2}$, an optimal orientation $\iota = 0$, and a constant ellipticity $\epsilon$. We note that the strain scales linearly with the moment of inertia, which may be a factor of a few larger than our fiducial value. In principle, we can choose to model the ellipticity as a function of time. However, over the long timescales considered here, the ellipticity is not expected to evolve significantly; the internal magnetic field that likely causes the stellar deformation gets wound up on the Alfv\'en timescale, which for these systems is $\ll 1$s \citep[e.g.,][]{Shapiro2000}. Although it is possible to have an evolution of the ellipticity through other mechanisms such as stellar cooling, the effect is similar to the angle between the star's principal moment of inertia and its rotation axis evolving due to, for example, the spin-flip instability (see Sec. \ref{sec:conclusion}). We leave this generalization for future work.
\subsection{Gravitational-wave energy budget}\label{subsec:Egwt}
We also consider the energy budget of the gravitational wave emission to determine allowed regions of the parameter space. The total power emitted in gravitational waves is 
\begin{equation}
\dot{E}_{\textrm{gw}}(t) = -\frac{32G}{5c^5}I_{zz}^{2}\epsilon^2\Omega^{6}(t).
\label{Eq. power}
\end{equation}
We substitute our gravitational-wave frequency evolution Eq. (\ref{Ch2: Eq.fgw(t)}) for the evolution of the star's angular frequency and integrate to determine the energy emitted in gravitational waves for a constant braking index 
\begin{equation}\label{Eq. Egw}
E_{\textrm{gw}}(t) = -\frac{32\pi^6G}{5c^5}I_{zz}^2\fgwo^6\epsilon^2\tau\frac{n - 1}{n - 7}\left[\left(1 + \frac{t}{\tau}\right)^{\frac{7 - n}{1 - n}} - 1 \right].
\end{equation}
This energy evolution is different to a standard continuous-wave signal as the strain evolves as a function of time. The total energy emitted in gravitational waves must be less than the initial rotational energy, $E_{\textrm{rot}}$ of the system
\begin{equation}
\label{eq. energy_budget}
\lvert E_{\textrm{gw}}(t)\rvert < E_{\textrm{rot}},
\end{equation}
where
\begin{equation}
E_{\textrm{rot}} = \frac{1}{2}I_{zz}\fgwo^2\pi^2.
\end{equation} 
We can use this condition to check if a given parameter space is physical. 
\begin{figure}[!htbp]
\centering
\includegraphics[width=0.5\textwidth]{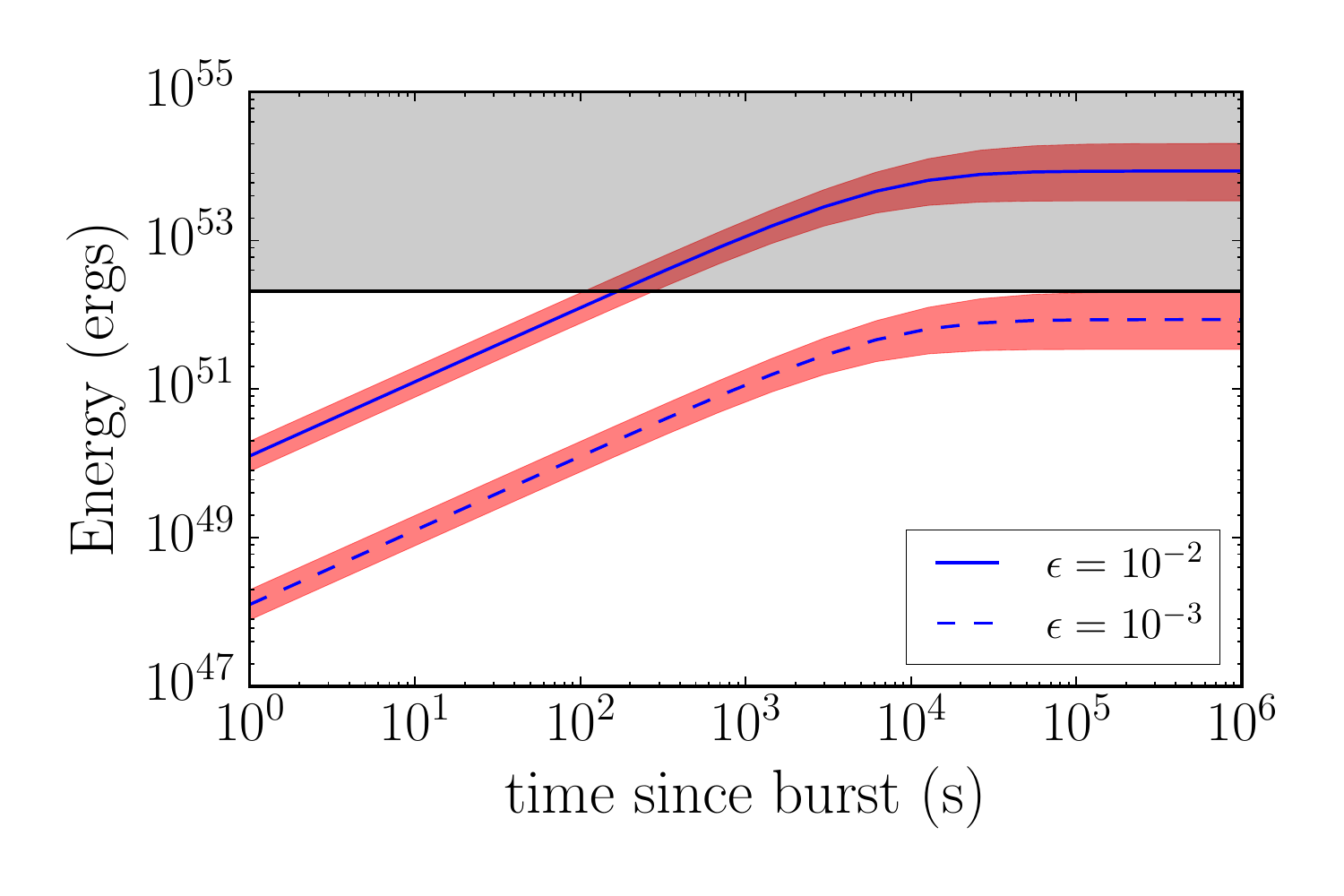}
\caption{The energy budget of a post-merger remnant inferred from GRB140903A with ellipticity $\epsilon = 10^{-2}$ (solid curves) and $10^{-3}$ (dashed curves) with the red shaded region indicating the 2$\sigma$ confidence interval. The grey shaded region above the solid black horizontal line is nonphysical as discussed in Sec. \ref{subsec:Egwt}.}
\label{fig:Energy Budget}
\end{figure}
Figure \ref{fig:Energy Budget} illustrates, for a post-merger remnant inferred from GRB140903A with a fiducial $I_{zz} = 10^{45}$ $\text{g cm}^{2}$, an ellipticity $\epsilon = 10^{-2}$ violates the energy-budget constraint. Based on these energy considerations the upper limit on ellipticity for GRB140903A is $\epsilon \approx 10^{-3}$. In reality, the moment of inertia for a long-lived post-merger remnant is likely higher than the fiducial value we use here, however all our limits can be scaled appropriately for different values of $I_{zz}$.  In particular, the moment of inertia is inversely proportional to the inferred upper limit on ellipticity, because the rotational energy grows linearly with $I_{zz}$, but the gravitational-wave energy grows quadratically.  Our fiducial moment of inertia therefore provides a conservative limit on the ellipticity. 
\subsection{Optimal matched filter statistic}\label{subsec:rho_opt}
\begin{figure*}[t]
  \begin{tabular}{cc}     
        \includegraphics[width=1.0\textwidth]{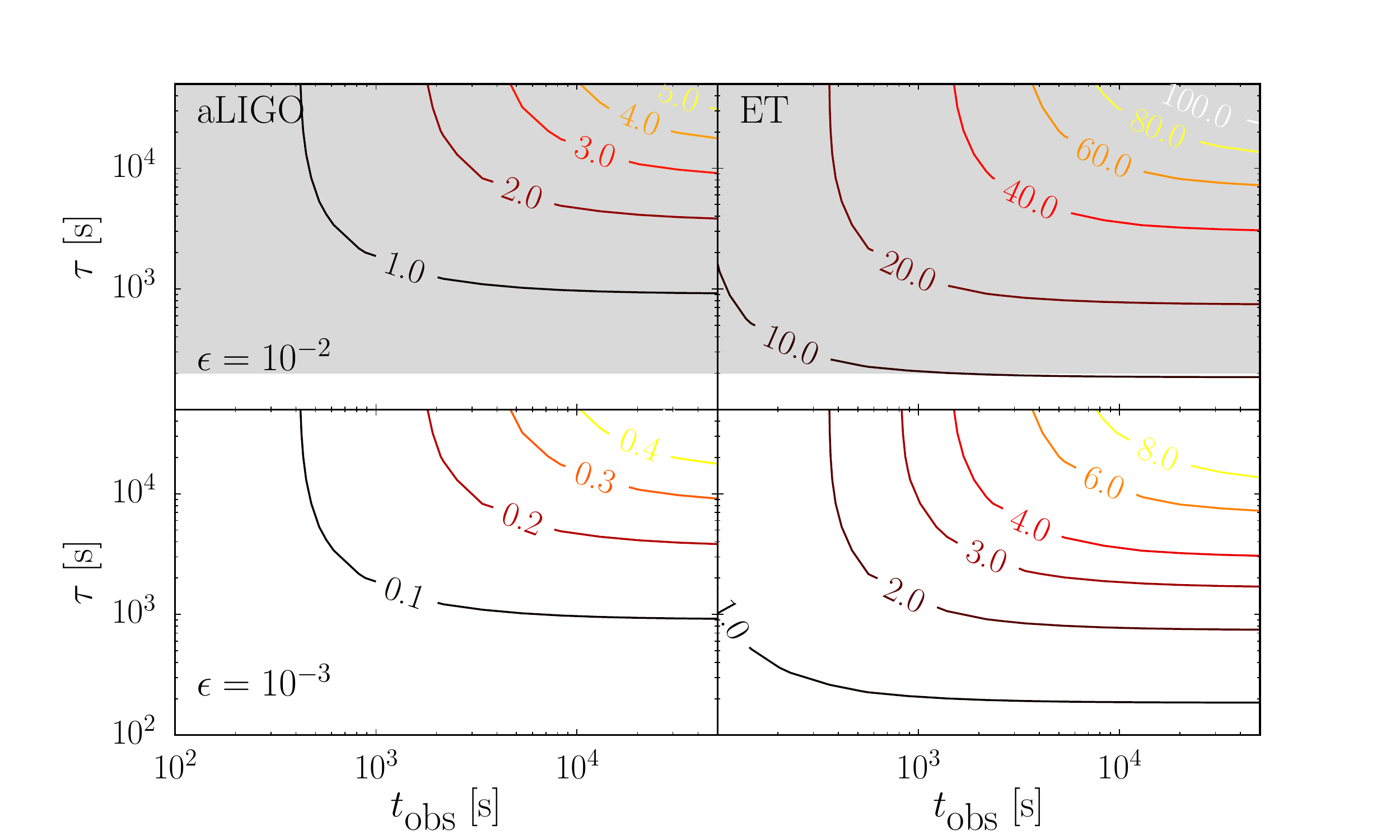} \hspace{-.4cm}
  \end{tabular}
  \caption{Optimal matched-filter signal-to-noise ratio $\rho_{\textrm{opt}}$ for a typical SGRB post-merger signal at a distance of $40$ Mpc as a function of the gravitational-wave observation time $t_{\textrm{obs}}$ and the spin-down timescale of the system. The left panels shows $\rho_{\textrm{opt}}$ for aLIGO with $\epsilon = 10^{-2}$ (top panel) and $\epsilon = 10^{-3}$ (bottom panel). The right panels show the same but for ET. The shaded region is nonphysical as the implied gravitational-wave energy emitted by the neutron star is greater than the available energy budget (see Sec. \ref{subsec:Egwt}). A $\rho_{\textrm{opt}} > 4.4$ is considered detectable.} 
\label{Fig. rho_opt}  
\end{figure*}

The matched-filter signal-to-noise ratio $\rho$ is given by \cite{Cutler1994}
\begin{equation}
\rho = \frac{\langle h|u\rangle}{\sqrt{\langle u|u\rangle}},
\label{eq:snr}
\end{equation}
where $h = s + n$ is the combination of signal $s$ and noise $n$, $u$ is the template, and $\langle a|b\rangle$ denotes the noise-weighted inner product \citep{Cutler1994}, defined by
\begin{equation}
\label{Ch3: Eq. inner_product}
\langle a|b\rangle =  4 \Re \int_{0}^{\infty} \frac{\tilde{a}^{\ast}(f)\tilde{b}(f)}{S_h(f)}df.
\end{equation}
Here $\tilde{a}$ denotes the Fourier transform of $a$, $\tilde{a}^\star$ its complex conjugate, and $S_{h}(f)$ is the noise power spectral density. 
The optimal matched-filter signal-to-noise ratio $\rho_{\textrm{opt}}$ is achieved when the template matches the data precisely:
\begin{equation}\label{eq:snr_exp}
\rho_{\textrm{opt}} = \sqrt{\langle h|h\rangle}.
\end{equation}
In this analysis, the threshold signal-to-noise ratio required to make a detection is $\rho_{\textrm{threshold}} = 4.4$, which is derived in Sec. \ref{sec:pipeline}.
In Fig. \ref{Fig. rho_opt} we show the region of parameter space where we could detect a signal from a post-merger remnant at the same distance as GW170817 ($40$ Mpc). 
We assume $I_{zz} = 10^{45}$ $\text{g cm}^{2}$, $\epsilon = 0.01$ (top panel) and $\epsilon = 0.001$ (bottom panel), $n = 2.71$ and $\fgwo = 2050$ Hz. We use these values of $\fgwo$ and $n$ as they are the maximum likelihood parameters from GRB140903A using the method detailed in Sec. \ref{sec:GRBafterglow}. 
The left-hand side of Fig. \ref{Fig. rho_opt} shows it is theoretically possible for gravitational waves from such an object to be observable by aLIGO operating at design sensitivity \cite{observing_aligo} if $\tau \gtrsim 4 \times 10^4$ s and $t_{\textrm{obs}} \gtrsim 4 \times10^4$ s. 
The right-hand side shows that the Einstein Telescope (ET), a proposed third generation detector \cite{ET}, can detect such a signal if $\tau \gtrsim 10^2$ s and $t_{\textrm{obs}} \gtrsim 10^2$ s for $\epsilon = 10^{-2}$.
We note that GRB140903A has $\tau = 17207 \pm 1880$ s. 
However, as shown in Sec. \ref{subsec:Egwt} this large ellipticity is nonphysical for GRB140903A-like post-merger remnant in all of the parameter space required to detect a signal with aLIGO. 
A physically realistic ellipticity $\epsilon = 10^{-3}$ rules out any prospect of detection with aLIGO for a GRB140903A-like post-merger signal at $40$ Mpc and requires $\tau \gtrsim 10^4$ s and $t_{\textrm{obs}} \gtrsim 10^4$ s for detecting the same signal with ET.

The optimal matched filter signal-to-noise ratio (Eq. \ref{eq:snr_exp}) can also be used to estimate the distance out to which we can detect a signal.
\begin{figure}[htbp]
\centering
\includegraphics[width=0.5\textwidth]{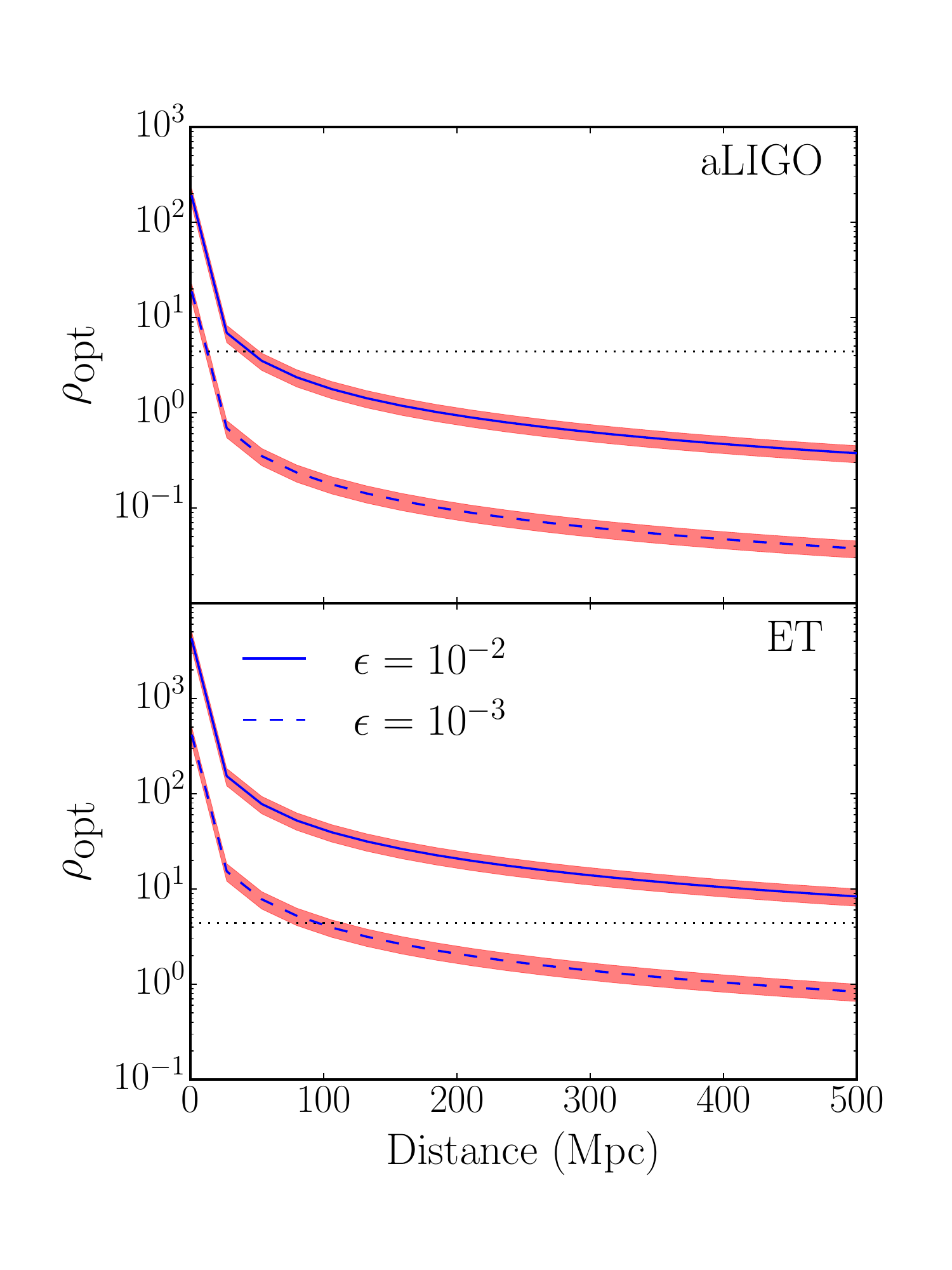}
\caption{The optimal matched-filter signal-to-noise ratio $\rho_{\textrm{opt}}$ as a function of distance for a millisecond magnetar inferred from GRB140903A for aLIGO (top panel) and ET (bottom panel) for two different ellipticities; $\epsilon = 10^{-2}$ (solid curves) and $\epsilon = 10^{-3}$ (dashed curves). The red shaded region indicates the $2\sigma$ confidence interval from the posteriors shown in Fig. \ref{fig:corner}. A threshold $\rho_{\textrm{opt}} = 4.4$ is indicated by a black horizontal dotted line. Any value above this threshold is detectable by aLIGO at design sensitivity. All curves are constructed using an observation time of $5\times10^4$ s.}
\label{fig. kappa GRB140903A}
\end{figure}
Figure \ref{fig. kappa GRB140903A} shows that with aLIGO at design sensitivity the furthest distance we can detect a signal with maximum likelihood parameters inferred from GRB140903A is $40$ and $4$ Mpc for $\epsilon = 10^{-2}$ and $10^{-3}$ respectively, while with ET the distances are $900$ Mpc and $90$ Mpc respectively. As we showed in Sec. \ref{subsec:Egwt}, for the parameters inferred from GRB140903A only an ellipticity $\epsilon \leq 10^{-3}$ is physical, post-merger remnants with longer spin-down timescale, $\tau$, can be detected to larger distances assuming that $\epsilon \sim 10^{-3}$ is physical for those parameters.

The optimal matched filter is the maximum signal-to-noise ratio one can achieve in a matched filter search. In practice, this limit is unobtainable with current computational resources. 
As shown by Fig. \ref{Fig. rho_opt} and Fig. \ref{fig. kappa GRB140903A}, to achieve $\rho_{\textrm{opt}} \geq 4.4$ and make a detection of gravitational waves, we need to observe a signal for at least $ \sim 10^4$ seconds with aLIGO at design sensitivity. 
At large observation times, the volume of parameter space imposed by uniform priors becomes unfeasible for a realistic gravitational-wave search (see Sec. \ref{sec:pipeline}). In the following section, we demonstrate how to constrain the priors, and hence the search parameter space, using X-ray observations of SGRBs
\section{\label{sec:GRBafterglow}X-ray afterglow}
\begin{figure}[!htbp]
\centering
\includegraphics[width=0.5\textwidth]{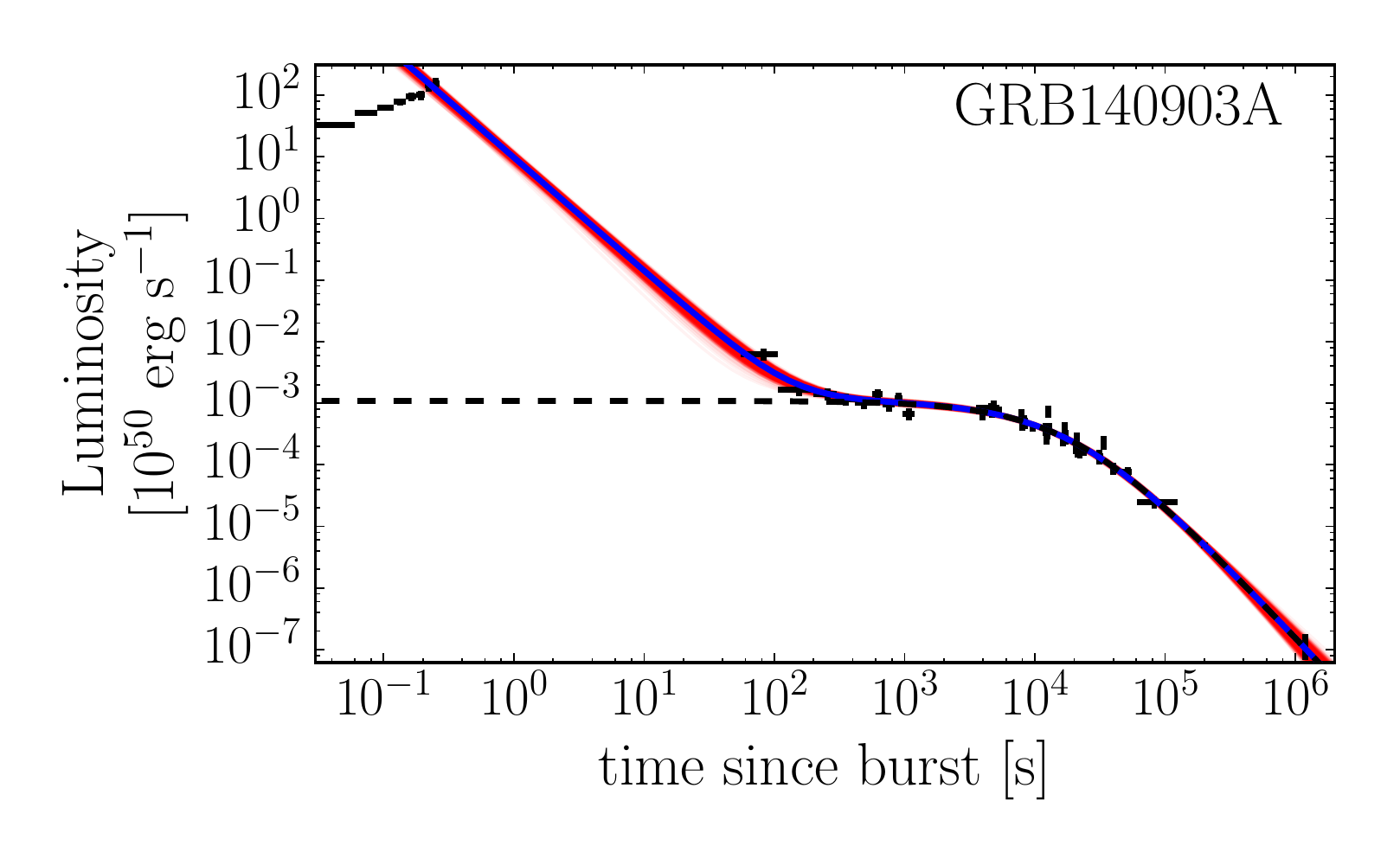}
\caption{$\gamma$ and X-ray lightcurves for GRB140903A. Black points are data from \textit{Swift} and Chandra satellites. The blue curve shows the maximum likelihood model described in Sec. \ref{sec:GRBafterglow}. The dark red band is the superposition of 800 models randomly drawn from the posterior distribution (shown in Fig. \ref{fig:corner}). The dashed black curve is the model for the luminosity from the nascent neutron star (Eq. \ref{eq:luminosity}).}
\label{fig:lightcurve}
\end{figure}

\begin{figure}[!htbp]
\centering
\includegraphics[width=0.5\textwidth]{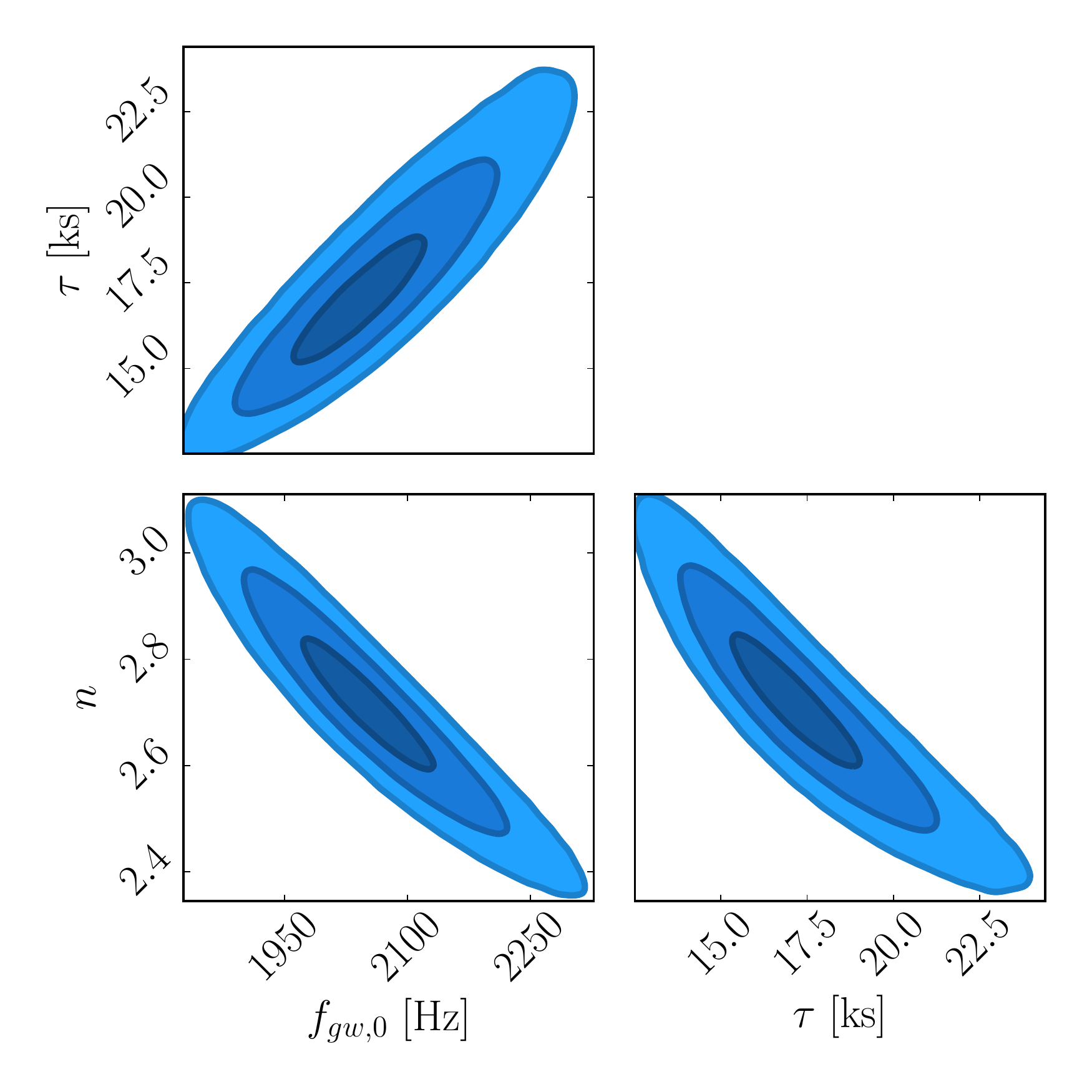}
\caption{Posterior distribution for $\fgwo$, $n$, and $\tau$ for GRB140903A. These posteriors are used as priors to build a GRB specific template bank. Shown are one-,two-, and three-sigma confidence levels. This figure is generated using the ChainConsumer software package \citep{Hinton2016}.}
\label{fig:corner}
\end{figure}

Short gamma-ray bursts are often followed by X-ray emission lasting up to many tens of thousands of seconds \cite{Rowlinson2010, Rowlinson2013, LaskyLeris2017, Lu2015}. Such an X-ray afterglow was not observed for GRB170817A. In Fig. \ref{fig:lightcurve} we show the X-ray afterglow of GRB140903A with data from the Neil Gehrels \textit{Swift} and Chandra satellites \citep{Troja2016}. 
\citet{Rowlinson2013} modelled the X-ray afterglows of several SGRBs with two components. Firstly, an initial power-law decay,
\begin{equation}
\label{Eq. Powerlaw}
L(t) = At^{-r},
\end{equation}
where $L$ is the luminosity, $A$ is the power-law amplitude, and $r$ is the power-law exponent. 
Here, the decay exponent can be fixed to $r = \Gamma_{\gamma} + 1$, where $\Gamma_{\gamma}$ is the photon index of the prompt emission, or allowed to vary. The second component is a luminosity law to model the energy injection from a millisecond magnetar that is spinning down through magnetic dipole radiation ($n = 3$) \cite{Zhang2001, Dai1998}. 
\citet{LaskyLeris2017} extended this model to include other forms of radiation causing spin-down, which is derived by utilising the general torque equation (Eq. \ref{Ch2: Eq. Torque}). The luminosity of the second component therefore comes directly from the nascent neutron star, and can be expressed as
\begin{equation}
\label{eq:luminosity}
L(t) = L_{0}\left(1+ \frac{t}{\tau}\right)^{\frac{1 + n}{1 - n}},
\end{equation}
where, $L_0$ is the initial luminosity at the onset of the plateau phase and is related to the initial gravitational-wave frequency $\fgwo$ by 
\begin{equation}
L_0 = \frac{\fgwo^2\pi^{2}I_{zz}\eta}{2\tau},
\end{equation}
where $\eta$ encodes the efficiency of converting spin-down energy to X-rays.
Our numerical model involves fitting Eq. (\ref{Eq. Powerlaw}) and (\ref{eq:luminosity}) to the X-ray observations from \textit{Swift} and Chandra. 
However, instead of fitting $L_0$ we fit our initial gravitational-wave frequency $\fgwo$. We use a Markov Chain Monte Carlo algorithm \cite{Foreman-Mackey2013} to fit the X-ray afterglow of SGRBs with our model using uniform priors for $f_{\text{gw}}$, $n$, $\tau$, $A$, and $r$ between [$\log_{10}(-1)$, $\log_{10}(5)$], [$\log_{10}(2)$, $\log_{10}(6)$], [$0$, $6$], [$\log_{10}(-10)$, $\log_{10}(5)$], and [$-2$, $5$] respectively.
Fits we have made to GRB140903A are shown in Fig. \ref{fig:lightcurve}. We determine the posterior distribution on our parameters $\fgwo$, $\tau$, and $n$ which are shown in Fig. \ref{fig:corner}. In the following section, we discuss how these posteriors can be used as priors for a targeted search for the post-merger remnant associated with an SGRB. 
\section{\label{sec:pipeline}Gravitational-wave search pipeline}
Here we describe a pipeline to search for gravitational waves from a spinning down millisecond magnetar. The algorithm can be summarised as follows:
\begin{enumerate}
    \item Generate posterior distributions on the three waveform parameters $\fgwo$, $n$ and $\tau$ using the X-ray afterglow observations of a specific SGRB as described in Sec. \ref{sec:GRBafterglow}.
    \item These posterior distributions, along with uniform priors on $\Phi_0$ and $\cos\iota \in [0,1]$, serve as priors for our waveform model. Template waveforms are generated from points in these priors.
    \item Templates are used to calculate the matched filter signal-to-noise ratio using LIGO data at the time of the SGRB.
\end{enumerate}
The same pipeline can also be adopted with unconstrained uniform priors in step 1, in the case where no X-ray data is available. However, the number of templates required for a matched-filter search becomes computationally unfeasible. We quantify this throughout this section. 

We calculate the fitting factor $FF$ \cite{Apostolatos1995}, also commonly referred to as the overlap \citep[e.g.,][]{Cornish2012}. The fitting factor is the penalty in signal-to-noise ratio one suffers due to comparing templates that do not precisely match the signal: $FF = \rho/\rho_{\textrm{opt}}$. 
We want to minimize this penalty while maximizing the signal-to-noise ratio.

To calculate the $FF$ we randomly draw one value of each parameter from our priors and construct a model waveform using the waveform model described in Sec \ref{sec:waveform}. 
We assume this is our true template, $h_{\textrm{T}}$. 
We determine the optimal matched filter signal-to-noise ratio for this template using Eq. (\ref{eq:snr_exp}),
We randomly draw from our priors excluding our `true template' and create a random template, $h_{\textit{i}}$, where $\textit{i}$ labels the $\textrm{i}^{\textrm{th}}$ drawn sample. We compute the matched filter signal-to-noise ratio (Eq. \ref{eq:snr}), $\rho_{\textit{i}}$. We calculate $\rho_{i}$ for N random templates.
In the limit of infinite templates, $\textrm{max}(\rho_{\textrm{i}}) \to \rho_{\textrm{opt}}$. 

The maximum fitting factor is defined as
\begin{equation}
\label{Ch3: Eq. FF}
FF = \frac{\text{max}(\rho_{\textrm{i}})}{\rho_{\textrm{opt}}},
\end{equation}
where $\text{max}(\rho_{\textrm{i}})$ is the maximum matched-filter signal-to-noise ratio from a population of N templates. In the limit of an infinite number of templates, $FF \to 1$, assuming our signal parameters are within our template parameter space. 
Creating a large number of templates is computationally expensive. We therefore want to minimise the number of templates we need. Additionally, we want to maximise our signal-to-noise ratio by creating templates for a longer duration.
\begin{figure*}[!htbp]
\begin{tabular}{cc!}
\includegraphics[width=0.9\textwidth]{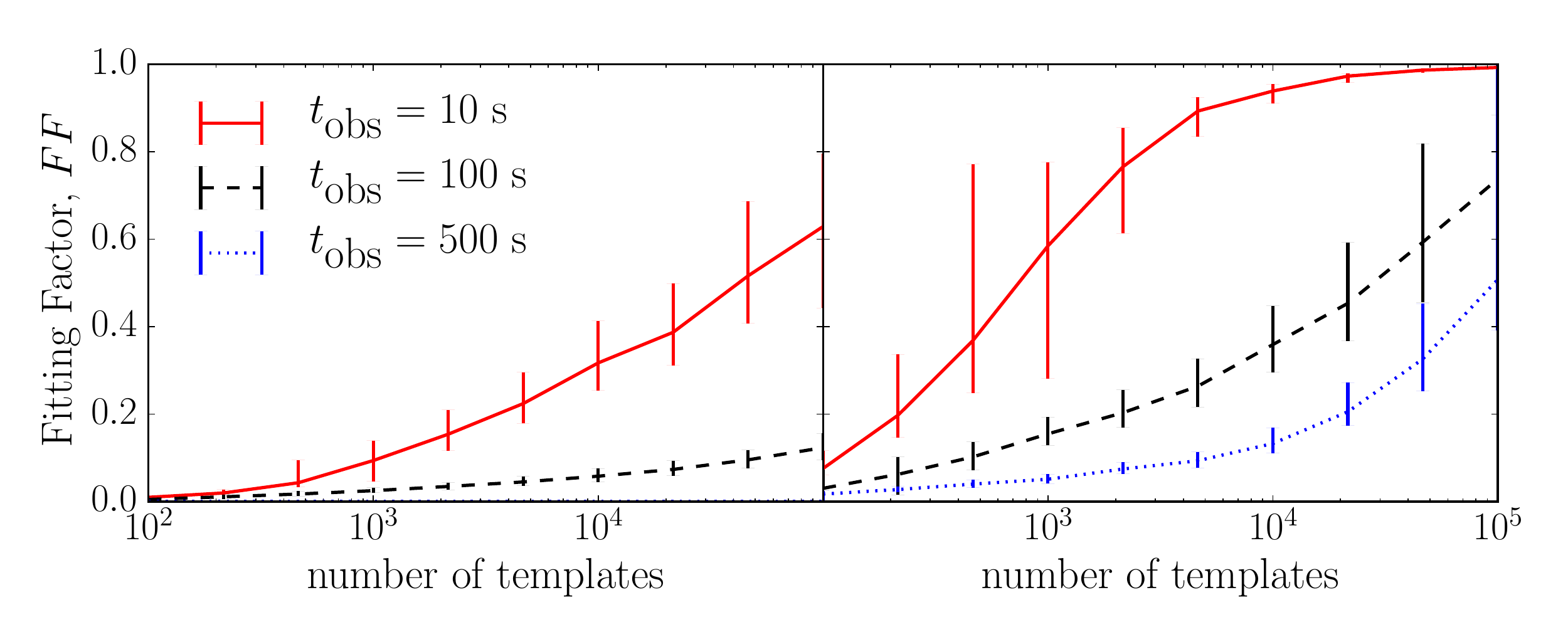}
\end{tabular}
\caption{The fitting factor ($FF$) as a function of the number of templates with unconstrained parameter priors (left panel) and priors constrained by X-ray afterglow observations (right panel) for observation times $t_{\textrm{obs}} = 10$ seconds (solid lines), $t_{\textrm{obs}} = 100$ seconds (dashed lines), and $t_{\textrm{obs}} = 500$ seconds (dotted lines). The error-bars indicate one sigma.}
\label{fig: FF}
\end{figure*}

In Fig. \ref{fig: FF} we show the scaling of $FF$ with the number of templates in the template bank for different $t_{\textrm{obs}}$ and two different priors: an unconstrained uniform prior (left panel) where the priors on $f_{\text{gw,0}},$ $n$ and $\tau$ are $[500, 3000]$ Hz, $[2.5, 5]$ and $[350, 35000]$ s, respectively, and the constrained posterior priors from using X-ray afterglow observations (right panel). The error bars indicate one sigma confidence levels, generated by repeating the analysis with 1000 different noise realizations. 

Figure \ref{fig: FF} shows that for $10^5$ templates, $FF = 0.62$ for $t_{\textrm{obs}} = 10$ s with uniform priors.
A fitting factor $FF = 0.62$ implies that we lose $38 \%$ of the optimal matched-filter signal-to-noise ratio when running a matched-filter search. 
This recovery percentage is even worse for longer observation times, with $t_{\textrm{obs}} = 100$ s having $FF = 0.12$ for uniform priors with $10^5$ templates, indicating we lose $88 \%$ of the optimal matched-filter signal-to-noise ratio. 
Although $FF$ scales up for an increasing number of templates, the amount of templates required to construct a search that could detect potential signals is unfeasible computationally for uniform priors. 
Furthermore, as shown in Sec. \ref{sec:waveform}, real astrophysical signals likely require $t_{\textrm{obs}} > 1000$ s, and $FF$ at these $t_{\textrm{obs}}$ is significantly worse. 
Fortunately, $FF$ is comparatively better for constrained priors (right panel). 
For example, for $t_{\textrm{obs}} = 100$ seconds with $10^5$ templates, $FF = 0.72$ with constrained priors as opposed to $0.12$ with uniform priors. 
In a real search we will likely require $t_{\textrm{obs}} > 10^3$ seconds and $10^6$ templates.
We have not calculated the $FF$ for these parameters as it is computationally expensive and requires an optimization step in the template generation to avoid using the high sampling frequencies throughout that are required at the beginning of the waveform. Furthermore, for aLIGO, detectable astrophysical signals require large $\tau$ values which are ruled out by the energy budget constraint; see Sec. \ref{subsec:Egwt}. In addition, constructing searches with observation times significantly larger than $\tau$ gives worse results as one no longer accumulates significant signal-to-noise for $t \gg \tau$.
Noting the scaling observed in $FF$, we expect $FF \approx  0.4$ for $t_{\textrm{obs}} = 10^4$ seconds with $10^6$ templates, an acceptable loss considering the gains from a longer signal duration.

We calibrate our pipeline by injecting signals into Gaussian noise coloured to match that of the expected strain sensitivity. This calibration is parameter dependent, so in a real search, we will need to do this for each SGRB.
We use the posteriors from GRB140903A to create a fake signal. In Sec. \ref{sec:waveform} we used the optimal matched filter signal-to-noise ratio (Eq. \ref{eq:snr_exp}) to determine an optimistic estimate for the distance out to which we can detect a signal (shown in Fig. \ref{fig. kappa GRB140903A}). These distances are optimistic, and as we quantified with $FF$, we suffer a loss in signal-to-noise due to having imperfect templates.

We define a horizon distance as the distance to which a detector with a given sensitivity can observe events with a given significance in a real matched-filter search. 
We start with the matched filter signal-to-noise ratio $\rho$ (Eq. \ref{eq:snr})
We determine a signal-to-noise ratio threshold $\rho_{\textrm{threshold}}$, which is the minimum signal-to-noise ratio to claim a detection with aLIGO at design sensitivity with a single detector. 
To determine this threshold, we calculate $\rho$ using Eq. (\ref{eq:snr}) with noise-only realisations ($s = 0$) and for \textit{N} templates. We take the maximum $\rho$ from \textit{N} templates and do this for multiple realisations of noise retaining the maximum $\rho$ each time. 
We determine the 99.7 percentile of our probability distribution on $\rho$ with no signal, which indicates that 99.7 \% of the time noise can mimic a signal (a false alarm).
Any detection needs $\rho > \rho_{\text{threshold}}$ to be significant. For our pipeline, the  $3\sigma$  $\rho_{\text{threshold}}$ is $4.4$ with $10^{4}$ templates and $1000$ realisations of noise, however the choice of this false-alarm rate is arbitrary.

We also establish a false dismissal probability, which quantifies when a real signal present in the data cannot be disassociated from the noise. As a result, it fails to be identified. 
To determine a horizon distance, we find the distance where our false dismissal probability is less than $10$ \%, which is done by repeating the procedure for determining $\rho_{\textrm{threshold}}$, but injecting signals at fixed distances.
We then determine at what distance less than $10$ \% signals have $\rho < \rho_{\textrm{threshold}}$.

Prior to this point, we have only considered a single detector; the signal-to-noise ratio grows approximately in quadrature for a network of $N$ similar detectors and therefore having an aLIGO-Virgo triple detector network will increase the horizon distance accordingly. In the future, with a network of 3G detectors such as ET and Cosmic Explorer, a similar increase in signal-to-noise ratio can be expected.
Other factors such as sky localization and time-varying $F_+$ and $F_\times$ will also affect the horizon distance. 
Considering these factors, in a real search we can expect our horizon distance for a GRB140903A inferred post-merger signal to be half the optimal matched-filter distance indicated by Fig. \ref{fig. kappa GRB140903A} as $\sim 2$ and $\sim 45$ Mpc for $\epsilon = 10^{-3}$ for aLIGO and ET respectively. 
\section{Conclusion} \label{sec:conclusion}
We have developed an algorithm to search for gravitational waves from a long-lived post-merger remnant of a binary neutron star merger. In Sec. \ref{sec:waveform}, we derive a waveform model for gravitational waves emitted from a spinning down millisecond magnetar. 
We detail and analyze a matched filter detection pipeline using this waveform model. We find that using X-ray observations from SGRB afterglows results in a significant decrease in parameter space resulting in a much improved and targeted search for a post-merger remnant. These X-ray guided priors can also be applied in other post-merger search pipelines.
Our analysis indicates for an ellipticity $\epsilon = 10^{-2}$ our pipeline can, in principle detect gravitational waves with aLIGO at design sensitivity out to $\sim 20$ Mpc for a fiducial moment of inertia $10^{45}$ g cm$^{2}$. If one ignores the energy-budget constraint, this fiducial value implies a conservative limit on the gravitational-wave strain and therefore horizon distance.  In reality, the moment of inertia of the remnant may be a factor few larger than this fiducial value; as the strain scales linearly with the moment of inertia, this implies the horizon distance may also be a factor of a few larger.  However, when including the energy-budget constraint, the horizon distance implied by a higher moment of inertia is lower due to the inverse relationship between the moment of inertia and the ellipticity.

It is the energy-budget constraint that ultimately sets the distance to which these post-merger remnants can be detected.  A large region of the parameter space is implausible, which lowers the horizon distance to $\sim 2$ Mpc for GRB140903A-like post-merger signals. The Einstein Telescope can detect a similar signal out to  $\sim 45$ Mpc. Post-merger signals with longer spin-down timescale $\tau$ will be detectable out to larger distances.

We are also investigating a more realistic model. The waveform model introduced here is simplified as the model assumes the neutron star is an orthogonal rotator. In this state, the principal eigenvector of the moment of inertia tensor is orthogonal to the star's rotation axis making the star an optimal emitter of gravitational waves. 
The neutron star is possibly driven to this orientation through the spin-flip instability \cite{Cutler2002, Mestel1981,1976Jones}, but the timescales involved are uncertain \cite{DallOsso2014, DallOsso2011, Lasky2016a}. As the system is driven to orthogonalization, it emits gravitational waves which we can include in our waveform model. We also have not accounted for time-varying $F_+$ and $F_\times$ terms.

Another extension is to constrain our parameter space further by including information obtained through parameter estimation on the binary neutron star inspiral gravitational-wave signal. 
Specifically, we can constrain the inclination of the source which should increase the pipeline sensitivity. The X-ray afterglow observations also suggest an evolution of the braking index with time with the system evolving from gravitational-wave dominated spin-down to magnetic dipole. This evolution of the braking index is something we can include in our model. 
\section{Acknowledgments}
We are grateful to David Keitel, Hou-Jun L\"u, and the anonymous referee for comments on the manuscript as well as the LIGO post-merger group for insightful discussions. NS is supported through an Australian Postgraduate Award. PDL is supported through Australian Research Council Future Fellowship FT160100112 and ARC Discovery Project DP180103155. 
\bibliographystyle{apsrev4-1} 
\bibliography{ref}
\end{document}